%% file: main.tex
\documentclass[fleqn,10pt]{JLA_article}
\usepackage[english]{babel}
\usepackage{booktabs}
\usepackage{multirow}
\usepackage{enumitem}
\usepackage{array}
\usepackage{tabularx}
\usepackage{float}
\usepackage{graphicx}
\usepackage{hyperref}
\usepackage{needspace}
\makeatletter
\patchcmd{\@maketitle}{\@Abstract\\[4pt]}{\@Abstract\ifhmode\\[4pt]\else\vspace{4pt}\fi}{}{\errmessage{Abstract spacing patch failed}}
\makeatother
\hypersetup{hidelinks,colorlinks,breaklinks=true,urlcolor=color2,citecolor=color1,linkcolor=color1,bookmarksopen=false,pdftitle={Supporting or Replacing Students' Reasoning? Cognitive Offloading in AI-Assisted Proof Learning},pdfauthor={Eason Chen, Sophia Judicke, Kayla Beigh, Xinyi Tang, Isabel Wang, Nina Yuan, Zimo Xiao, Chuangji Li, Shizhuo Li, Reed Luttmer, Shreya Singh, Maria Yampolsky, Naman Parikh, Yvonne Zhao, Meiyi Chen, Scarlett Huang, Anishka Mohanty, Gregory Johnson, John Mackey, Jionghao Lin, Ken Koedinger}}

\PaperTitle{Supporting or Replacing Students' Reasoning?\\Cognitive Offloading in AI-Assisted Proof Learning}

\Authors{Eason Chen\textsuperscript{1}, Sophia Judicke\textsuperscript{1}, Kayla Beigh\textsuperscript{1}, Xinyi Tang\textsuperscript{1}, Isabel Wang\textsuperscript{1}, Nina Yuan\textsuperscript{1}, Zimo Xiao\textsuperscript{1}, Chuangji Li\textsuperscript{1}, Shizhuo Li\textsuperscript{1}, Reed Luttmer\textsuperscript{1}, Shreya Singh\textsuperscript{1}, Maria Yampolsky\textsuperscript{1}, Naman Parikh\textsuperscript{1}, Yvonne Zhao\textsuperscript{1}, Meiyi Chen\textsuperscript{1}, Scarlett Huang\textsuperscript{1}, Anishka Mohanty\textsuperscript{1}, Gregory Johnson\textsuperscript{1}, John Mackey\textsuperscript{1}, Jionghao Lin\textsuperscript{2}, Ken Koedinger\textsuperscript{1}}
\affiliation{\textsuperscript{1}Carnegie Mellon University, Pittsburgh, PA, United States.}
\affiliation{\textsuperscript{2}The University of Hong Kong, Hong Kong.}
\affiliation{Correspondence: \href{mailto:EasonC13@cmu.edu}{EasonC13@cmu.edu}; \href{mailto:jionghao@hku.hk}{jionghao@hku.hk}.}

\Notesname{Notes for Practice}
\note{Students assigned earlier access to AI support earned higher homework scores, but we found no significant advantage on the subsequent exam, which students completed without AI. Evaluations of AI tutoring should therefore include both AI-supported work and subsequent independent performance.}
\note{Students with lower baseline self-efficacy used the AI chatbot more often, and greater chatbot use was associated with lower subsequent exam scores. This pattern raises concern about cognitive offloading, in which students rely on AI to supply reasoning they would otherwise need to develop themselves. Instructors should ask students to explain AI-assisted solutions in their own words.}
\note{We recommend against using an open-ended AI chatbot as a tutor. AI learning tools should engage students in developing and explaining their own reasoning rather than simply supplying answers. Instructors and designers should explore tools that ask students to make an initial attempt and provide feedback on that work.}

\Abstract{We studied 153 students in an undergraduate proof-based mathematics course with staggered-access to an AI support system.
Students were randomly assigned to receive an AI chatbot and a homework-feedback interface earlier or later in the course.
During the period when only the early-access group had access, their homework scores were 2.56 points higher, but exam performance did not differ significantly between groups, suggesting that improved AI-supported homework performance did not translate into detectable gains on a subsequent exam.
After all students received access, every additional 100 messages sent to the AI chatbot was associated with a 3.02-point lower score on the subsequent exam. A mediation analysis showed a significant indirect association linking lower baseline self-efficacy to greater chatbot use and, in turn, lower exam scores: a one-point decrease in self-efficacy corresponded to 0.62 fewer exam points through chatbot use. These findings raise concerns that students may use the chatbot to offload their cognitive reasoning.
Use of the homework-feedback interface, which required students to submit their proof attempts before receiving feedback, was not negatively associated with exam performance. We recommend against using a chatbot for AI tutoring and suggest exploring a tutor embedded in the course activities.

}

\begin{document}
\flushbottom
\maketitle
\thispagestyle{fancy}

\section{Introduction}

Generative artificial intelligence (GenAI) can help students complete difficult coursework, but success with assistance does not necessarily mean that students can complete similar work independently. This distinction is especially important in proof writing, where students must construct and explain a chain of reasoning rather than only recognize a plausible solution \cite{weber2001student}. Explaining one's own reasoning can support learning \cite{chi1994eliciting}, and proof-specific self-explanation training can improve comprehension \cite{hodds2014self}. An evaluation of GenAI support should therefore examine both the work students complete with assistance and what they can later do without that assistance.

\begin{figure}[!ht]
\centering
\includegraphics[width=.90\linewidth]{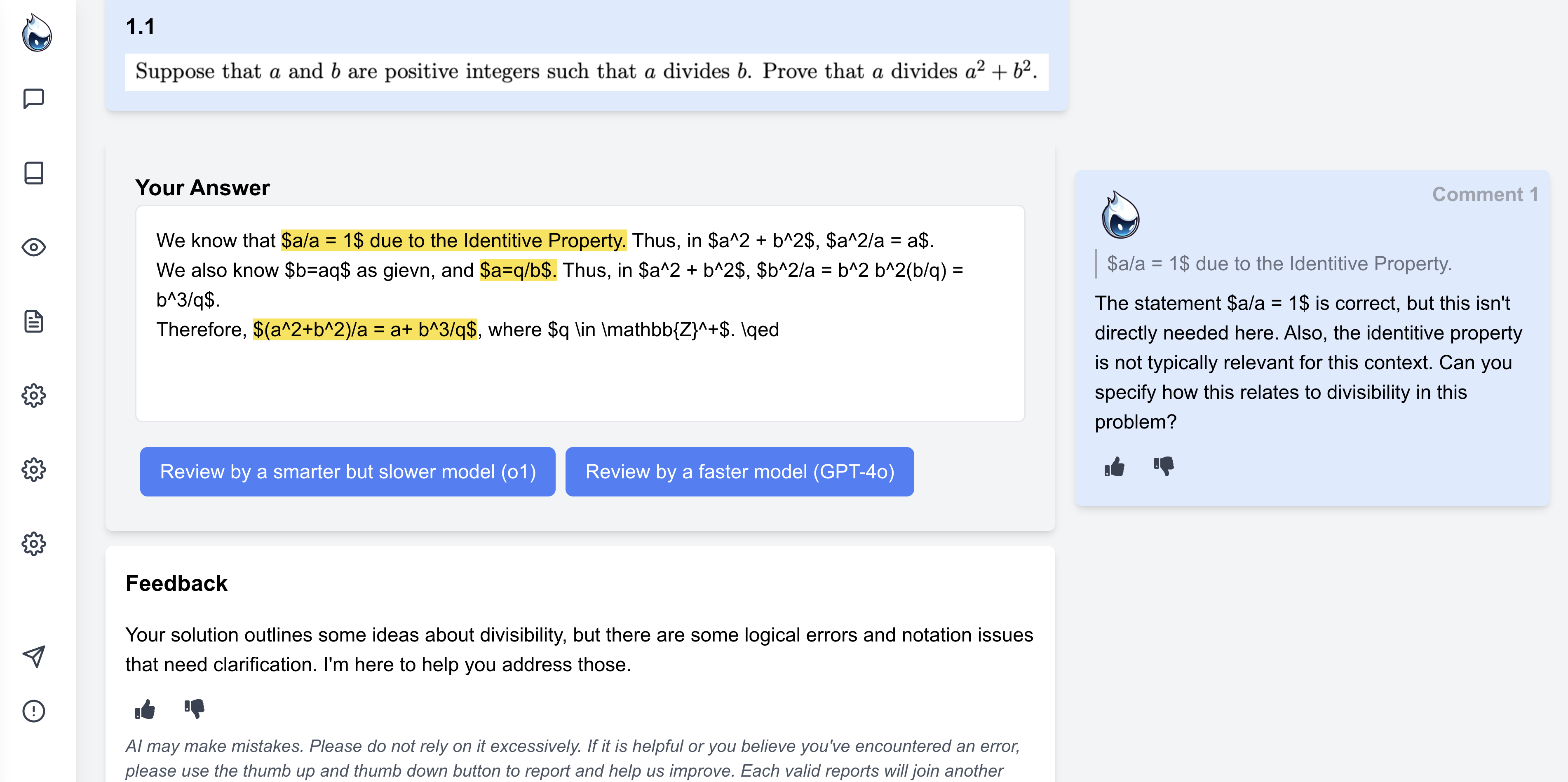}
\caption{Homework-feedback interface (Proof-Review-GPTutor), which provides feedback on a student's submitted proof attempt. The homework problem appears at the top, and the student's proof appears on the left. Highlighted passages link to localized comments on the right, while the overall feedback below summarizes issues across the submission. In this example, the feedback identifies logical and notational problems and asks the student to clarify how a highlighted statement supports the divisibility argument. The homework-feedback interface comments on the student's submitted proof rather than responding to an open-ended AI chatbot request.}
\label{fig:proof_review}
\medskip
\includegraphics[width=.96\linewidth]{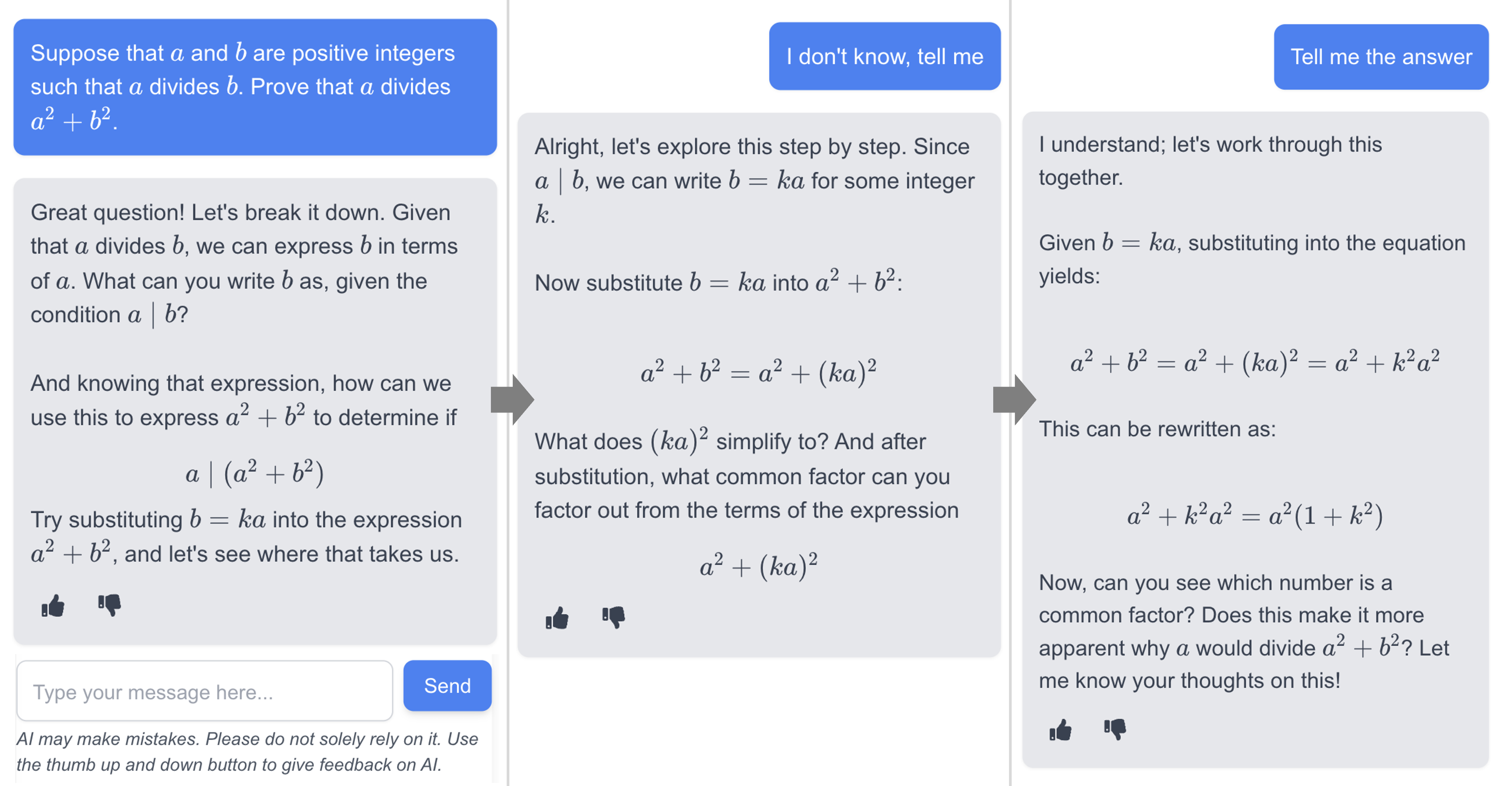}
\caption{AI chatbot interaction in which a student moves from receiving guidance to requesting a direct solution. Left: The student submits a divisibility proof problem as an answer-seeking behavior, and the AI chatbot responds with guiding questions and a suggested substitution. Middle: The student replies, ``I don't know, tell me,'' and the AI chatbot supplies the substitution step while asking the student to simplify and factor. Right: The student explicitly escalates their requests, ``Tell me the answer,'' and the AI chatbot supplies the factorization while still prompting the student to explain why it establishes divisibility. The student repeatedly requests more direct help instead of answering the guiding questions. The AI chatbot supplies increasingly explicit proof steps while continuing to ask the student to explain the reasoning.}
\label{fig:chat_sequence}
\end{figure}

Our AI support system provided two interfaces. The AI chatbot allowed students to ask mathematics questions without first submitting a proof attempt. Although its system prompt instructed it to avoid final answers and complete proof steps, students could still request them and press for more direct help. In contrast, the homework-feedback interface required students to submit a written proof before receiving critique. These differences may affect whether students use AI to develop their own reasoning or to obtain reasoning they have not attempted. Students received access to the two interfaces together and later chose which one to use. Figures~\ref{fig:proof_review} and~\ref{fig:chat_sequence} show the two interfaces.

Prior research gives several reasons to separate availability, use, and learning. The assistance dilemma holds that too little help can leave a student stuck, while too much can replace activity needed for learning \cite{koedinger2007exploring}. Cognitive offloading can reduce a task's processing demands \cite{risko2016cognitive}; in this setting, a student might use AI to obtain proof steps rather than generate them. Whether offloading supports learning depends on what reasoning the student performs with the assistance. In one mathematics field experiment, an unguarded chatbot improved performance while support was available but was followed by lower scores on an assessment without AI; tutoring guardrails reduced that pattern \cite{bastani2025generative}. Other recent work shows that automated feedback can improve proof submissions across repeated attempts, although that study did not assess later independent performance \cite{zhao2025autograding}.

Students who are less confident or already struggling may also choose to seek more help \cite{aleven2003help,bandura1997self,zimmerman2000self}. Understanding the relationship between AI use and later performance therefore requires examining both who seeks help and whether students use that help to develop or replace their own proof construction.

We studied these questions in an undergraduate proof-based mathematics course covering logic, sets, relations, and proof writing. Students were randomly assigned to receive the two interfaces after either Midterm~1 or Midterm~2. During this staggered-access period, only the early-access group could use the two interfaces for homework. Students completed the next exam independently without AI. We compared students assigned earlier and later access on the homework completed during this period and on the next exam.

After Midterm~2, all students could use the two interfaces, and usage was self-selected. We first examined which measured student characteristics were associated with using each interface more often and how usage counts related to Midterm~3. We then used mediation analysis to examine indirect associations between baseline self-efficacy and Midterm~3 through Midterm~2 performance and subsequent use of the two interfaces.

In this study, we ask three research questions (RQs):

\begin{itemize}[leftmargin=*]
    \item \textbf{RQ1:} How did assignment to earlier access to the two interfaces affect homework scores and scores on the next exam, which students completed independently without AI?
    \item \textbf{RQ2:} After all students received access, how was students' self-efficacy associated with the use of each interface? How were the number of student messages sent to the AI chatbot and the number of homework-feedback requests associated with scores on the next exam?
    \item \textbf{RQ3:} How was baseline self-efficacy indirectly associated with Midterm~3 scores through Midterm~2 performance and subsequent use of the two interfaces?
\end{itemize}

Our findings connect access to AI support with how students chose to use it. Students assigned earlier access earned higher homework scores, but we found no significant difference between access groups on the next exam. After all students received access, those with lower self-efficacy used the two interfaces more often, and students who sent more messages to the AI chatbot earned lower exam scores. The mediation analysis linked lower self-efficacy to greater AI chatbot use and, in turn, lower subsequent exam scores. Together, these findings suggest that AI may help students complete homework without strengthening their ability to construct proofs independently.

To support further study, we have also prepared annotation prompts, coding definitions, analysis scripts, and aggregate results that researchers can inspect and adapt to their own approved student--AI conversation datasets.

\section{Related Work}

\subsection{Learning Analytics for Instructional Decision-Making}

In the cycle described by \citeauthor{clow2012cycle}~\citeyear{clow2012cycle}, analysts derive metrics from learner data, educators use them to select interventions, and subsequent evidence is used to evaluate those interventions. \shortciteauthor{yan2024genaila}~\citeyear{yan2024genaila} apply this sequence to the opportunities and risks of generative AI. To interpret an activity count, researchers need a theoretical account of the behavior and an understanding of the course design. They also need to evaluate any instructional action based on the count. \citeauthor{winne2020construct}~\citeyear{winne2020construct} makes the same link explicit through construct and consequential validity: theory must justify what a trace represents and whether actions based on it have the intended consequences. \shortciteauthor{gasevic2015learning}~\citeyear{gasevic2015learning} similarly argue that analytics should remain anchored in learning rather than in computational prediction alone.

Instructors need course context to decide how to respond to activity records. The situated model proposed by \citeauthor{wise2019teaching}~\citeyear{wise2019teaching} describes four activities: asking questions, interpreting patterns in the course context, deciding whether to intervene, and evaluating the effects. In that study, instructors rarely evaluated the effects. The intervention framework proposed by \citeauthor{wise2014interventions}~\citeyear{wise2014interventions} further emphasizes that analytics need to be integrated into teaching activities to support their productive use. For AI-supported proof work, this means using repeated requests to identify where a student needs help and reviewing the student's reasoning before choosing an instructional response.
When analytics inform decisions about students, educators should review the evidence in context, prioritize support over punishment, and allow interpretations to be revised \cite{slade2013ethical}.

\subsection{AI Support, Help-Seeking, and Independent Reasoning}

LLM-powered course assistants can make educational support flexible and available on demand, but outcomes depend on the form and timing of support \cite{liu2025understanding,bastani2025generative}. Recent mathematics studies report gains on supported work under different safeguards and interface designs \cite{bastani2025generative,zhao2025autograding}; only the study by \citeauthor{bastani2025generative} also assessed later performance without AI. \citeauthor{koedinger2007exploring}~\citeyear{koedinger2007exploring} frame this tradeoff as the assistance dilemma: too little help can leave students stuck, while too much can replace the cognitive activity required for learning. Open-ended chat increases flexibility and learner initiative, but it can also make complete solutions easier to request. Feedback on submitted homework requires an attempt before critique and centers the exchange on the student's own reasoning.

In a randomized programming experiment, \shortciteauthor{bassner2026stress}~\citeyear{bassner2026stress} compared the hint-based Iris tutor, unrestricted ChatGPT, and no AI during a 90-minute exercise. Both AI groups earned higher exercise scores, but neither showed a significant advantage in pre--post knowledge gains or code comprehension. We examine a related question during ordinary course work: whether earlier AI access improves homework scores and performance on a subsequent exam without AI.

\shortciteauthor{schemmer2023appropriate}~\citeyear{schemmer2023appropriate} define advice-level appropriate reliance as following correct AI advice while rejecting incorrect advice; \shortciteauthor{vasconcelos2023explanations}~\citeyear{vasconcelos2023explanations} show that explanations and verification costs can shape over-reliance. In proof learning, using AI advice also raises a question about what students can subsequently construct for themselves. Reading a plausible proof differs from constructing one, so students still need to explain their own reasoning and receive feedback on it \cite{chi1994eliciting,hodds2014self,aleven2002effective}. We therefore examine how the frequency of AI use relates to students' subsequent ability to construct proofs without assistance.

Students may seek help for different reasons. \shortciteauthor{aleven2003help}~\citeyear{aleven2003help} show that help use in interactive learning environments reflects learner knowledge, metacognition, motivation, and system design; it can be effective, ineffective, or avoided even when needed. A negative usage--outcome association may therefore partly reflect struggling students choosing to seek more help rather than harm from assistance. More generally, self-efficacy is associated with effort and persistence and interacts with self-regulated learning \cite{bandura1997self,schunk1991self,zimmerman2000self}. We therefore first model usage counts as an outcome and then examine later performance with prior achievement and self-efficacy as covariates.

A recent programming-course study illustrates why the source of usage measures also matters. \citeauthor{lepp2025generative}~\citeyear{lepp2025generative} found that students who reported more frequent AI use earned lower programming-test and exam scores in a course that allowed AI for homework but required those assessments to be completed without AI. Their measures came from a mid-course survey. We instead use logged messages and homework-feedback requests and account for prior scores and self-efficacy when examining subsequent exam performance.

\subsection{Chatbot Dialogue Analysis and Learning Outcomes}

Recent learning-analytics studies demonstrate the value of analyzing dialogue content rather than treating chatbot use as a single count. \shortciteauthor{lai2025processaction}~\citeyear{lai2025processaction} found that higher-scoring students engaged more frequently in reflective and evaluative activities during Socratic-chatbot conversations, whereas lower-scoring students focused more on searching for answers. \shortciteauthor{mcnichols2026studychat}~\citeyear{mcnichols2026studychat} annotated StudyChat interactions, evaluated their annotator against a held-out human-coded set, and related dialogue acts to assignment and LLM-free exam outcomes. We use dialogue examples to interpret the associations among self-efficacy, logged AI chatbot use, and subsequent exam performance.

A preprint by \shortciteauthor{rinja2026unpacking}~\citeyear{rinja2026unpacking} reanalyzes the same StudyChat data through sequences of student requests and AI responses, rather than providing an independent replication. Larger shares of AI explanations were associated with higher average grades and larger shares of pasted assignment text with lower grades, although their broader model of help-seeking strategies and AI roles found no significant predictors of grades. This distinction motivates examining what follows a request.

Content labels may make aggregate counts more interpretable, but \shortciteauthor{misiejuk2024augmenting}~\citeyear{misiejuk2024augmenting} caution that LLM coding of educational discourse can be unreliable for nuanced, context-dependent constructs and requires task-specific validation. We distinguish agreement between human coders from agreement between the classifier and the human consensus.

\section{Course and the AI Support System}

The study took place in an undergraduate math course at Carnegie Mellon University. The course covered logic, sets, relations, and proof writing. Students submitted homework through ordinary course procedures, and course policy required students to complete exams independently without AI. The AI support system was GPTutor.

The AI support system provided the two interfaces. The \emph{homework-feedback interface} required a student to submit a written proof attempt. It returned overall feedback and comments tied to specific parts of the proof. Instructors provided the homework prompt and a reference solution as context, and the instructions asked the model to critique the submitted proof rather than begin an open conversation. Figure~\ref{fig:proof_review} shows the interface.

The \emph{AI chatbot} accepted mathematics questions without requiring a proof attempt and allowed repeated exchanges. Its system prompt instructed the model to avoid final answers and complete proof steps and instead use Socratic questions, scaffolds, and partial hints. Students could ask for concepts, procedures, next steps, or direct answers, and they could respond to hints or press for more direct help. Figure~\ref{fig:chat_sequence} shows a de-identified three-panel example. The example shows the student responding to guidance with increasingly direct requests for a solution.

The two interfaces called the OpenAI application programming interface (API) without fine-tuning. The AI chatbot used the fixed \texttt{gpt-}\allowbreak\texttt{4o-}\allowbreak\texttt{2024-}\allowbreak\texttt{08-}\allowbreak\texttt{06} snapshot. The homework-feedback interface offered that snapshot and the slower \texttt{o1-}\allowbreak\texttt{preview-}\allowbreak\texttt{2024-}\allowbreak\texttt{09-}\allowbreak\texttt{12} model, with the same prompt, instructor reference solution, and structured feedback format. A different model labeled AI chatbot sessions after the course for the analysis below; students did not interact with that classifier.

The AI support system recorded timestamps, requests, sessions, and messages. Students submitted a proof before requesting homework feedback, whereas student messages sent to the AI chatbot could request different forms of help.

\section{Methods}

\begin{figure}[b!]
\centering
\includegraphics[width=\linewidth]{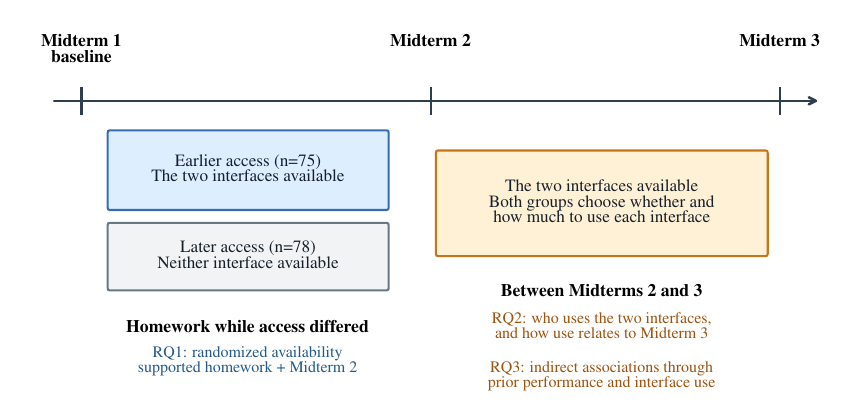}
\caption{Study design and analysis windows. After Midterm~1, students were randomly assigned earlier or later access to the two interfaces. Only the early-access group could use the two interfaces for homework during the early-access period. After Midterm~2, both groups could use the two interfaces. RQ2 examines which student characteristics predict use of the AI support system after all students received access, and how use of the AI support system relates to Midterm~3 scores. RQ3 examines indirect associations between baseline self-efficacy and Midterm~3 through Midterm~2 and subsequent interface use. Randomization applies to access timing, not interface choice or amount of use.}
\label{fig:study_design}
\end{figure}

We conducted three analyses that correspond to the research questions. First, we compared homework completed during the early-access period and Midterm~2 scores. Second, after all students had access, we examined who used each interface and whether frequency of use was associated with Midterm~3 scores. Third, we examined indirect associations between baseline self-efficacy and Midterm~3 through prior performance and interface use. Grades came from course records, usage counts came from system logs, self-efficacy came from a survey, and intent labels came from human coders and a separate classifier.

\subsection{Participants and Study Design}

The institutional review board (IRB) approved the study. The course enrolled 308 students. Homework and exams followed standard course procedures, and students were not permitted to use AI during exams.

After Midterm~1, students were invited to consent and complete a baseline survey. Participation was voluntary and did not affect grades. Participants were randomly assigned as individuals to early or later access. The early-access group received the two interfaces after Midterm~1 and Homework~4. The later-access group received them after Midterm~2. Therefore, only the early-access group could use the two interfaces for homework during this period. We selected the period between Midterms~2 and 3 for the usage analyses because both groups had access to the system, providing a larger sample for examining AI use than the earlier period, when only the early-access group had access. The system was actively maintained throughout this period.

Of 155 students who consented, we had grade and usage records for 153: 75 in the early-access group and 78 in the later-access group. These records were unavailable for the other two students. One student had no Midterm~2 score because they withdrew from the course. Four additional students had no Midterm~3 score for the same reason. No students withdrew from the research study. We retained available grades and usage records for students who withdrew from the course, and each model included students with data for that model's outcome and predictors. The homework model included all 153 students, the Midterm~2 model and models of counts of interface use included 152, and both the model relating usage counts to Midterm~3 and the mediation analysis included 148.
The course included a final exam after Midterm~3. The AI support system remained available, but data collection for this study ended at Midterm~3 because of budget constraints. The final exam was therefore not included in the analyses.

\subsection{AI Use for Homework and Study}

Students submitted homework in \LaTeX{} and were encouraged, but not required, to paste drafts into the homework-feedback interface before submission. They could use the AI chatbot for mathematics questions while doing homework or studying. Course policies did not restrict other external resources outside exams.

\subsection{Measures and Data Sources}

\noindent\textbf{Academic Performance.}
For the staggered-access analysis, we analyzed scores on Homework~5--6, which students completed while only the early-access group could use the two interfaces. This window was defined by the access schedule. The Homework~1--4 average measured prior homework performance, and Midterm~2 was the next exam completed without AI. Later usage analyses used Midterm~3 as the outcome, with Midterm~2 measuring prior performance. Exams assessed proof reasoning using problems similar in type to homework, but not the same questions. Course staff graded homework and exams through ordinary procedures. Bonus items allowed some standardized scores to exceed 100.

\noindent\textbf{System Interaction Logs.}\label{sec:logs}
The AI support system logged homework-feedback requests and AI chatbot interactions. For each participant, we counted homework-feedback requests and student messages sent to the AI chatbot between Midterms~2 and 3. The final labeling dataset contained 659 sessions with messages from 106 users. These sessions provided the corpus for intent labeling and the illustrative dialogue analysis.

\noindent\textbf{Labeling Student Intent in Chatbot Requests.}
Two undergraduate coders independently labeled student intent in a random sample of 80 sessions as help-seeking (requests for explanations or guidance), answer-seeking (requests for a complete answer or solution), or other. They agreed on 73/80 sessions (Cohen's $\kappa=.780$) before a doctoral researcher resolved disagreements. A separate GPT-5.2 classifier (snapshot \texttt{gpt-5.2-}\allowbreak\texttt{2025-12-11}, temperature 0) labeled intent from the first three messages of each session. On the same 80 sessions used to develop its prompt, classifier agreement with the consensus intent labels was $\kappa=.84$.

We compared two prompt versions applied to the same 659 sessions. They differed only in the definition of the other/meta category. Comparing them tests whether a small wording change altered who received the answer-seeking label. Both prompts classified 51 of 106 users as having at least one answer-seeking session, but assigned different classifications to six users.

Answer-seeking meant asking for an answer that could be submitted as coursework or a full solution, such as asking the AI chatbot to prove a statement without offering an attempt. Help-seeking meant asking to understand or improve one's own reasoning, such as requesting a hint or checking a step. Other/meta covered system questions, greetings, and off-topic messages; unclear messages could receive a separate unclear label. The classifier used the first three messages as context, but the intent label described only the first student message. A student could begin by seeking help and later request an answer, or begin by requesting an answer and later engage with the explanation.

\noindent\textbf{Dialogue Examples and Student Interviews.}
We selected two four-turn student--AI chatbot sessions to illustrate contrasting initial requests and how the exchanges developed. We also conducted semi-structured interviews with a convenience sample of 19 students from the course who volunteered to discuss their experiences with the AI support system, including their homework routines, understanding of course material, confidence, and use of other learning resources. We use excerpts from the interviews to help interpret the quantitative findings and understand how students used the AI support system. Bracketed ellipses indicate omitted text.

\noindent\textbf{Baseline Self-Efficacy.}
At baseline (after Midterm~1), participants
completed four items adapted from the nine-item Self-Efficacy scale reported by \citeauthor{pintrich1990motivational}~\citeyear{pintrich1990motivational}. The four items were: ``I'm certain I can understand the ideas taught in this course.''; ``I expect to do very well in this class.''; ``I am sure I can do an excellent job on the problems and tasks assigned for this class.''; and ``I think I will receive a good grade in this class.'' Items used a 1--7 response scale and were averaged so that higher values indicated greater confidence. Internal consistency was high (Cronbach's $\alpha = 0.91$).

\subsection{Resources for Reusing the Analysis}

The analysis code, annotation prompts, model settings, and aggregate results are available in our supplementary repository: \url{https://anonymous.4open.science/r/lak27-anon}.

\subsection{Statistical Analyses}

We analyzed the study data in Python. Each model included students with complete data for that outcome and its predictors, and we report model-based $p$ values. Later usage was counted only between Midterms~2 and 3. The association between student messages sent to the AI chatbot and Midterm~3 is the main usage result. The same model also estimates the association between homework-feedback requests and exam scores. The access and usage--exam ordinary least-squares models used heteroskedasticity-consistent standard errors (HC3), which allow error variance to differ across students. The mediation analysis used conventional OLS path estimates and bootstrap confidence intervals for indirect associations.

\noindent\textbf{Access, homework, and the next exam.}
For homework during the early-access period, we compared students according to their randomized access assignment, whether or not they actually used either interface. A maximum-likelihood mixed regression included a random intercept for each student, early versus later access, homework assignment, and the Homework~1--4 average. For Midterm~2, ordinary least squares (OLS) with HC3 standard errors compared access groups while controlling for Midterm~1.

\noindent\textbf{Student Characteristics and Frequency of Interface Use.}
We used negative-binomial regressions with log links and robust sandwich standard errors because usage counts included many zeros and varied widely. Each model included baseline self-efficacy and Midterm~2 scores. We report percentage differences in expected counts. These models include both users and nonusers, rather than comparing only students who used either interface.

\noindent\textbf{Interface Use and Midterm~3 Scores.}
We modeled Midterm~3 from the number of student messages sent to the AI chatbot and the number of homework-feedback requests while controlling for self-efficacy and Midterm~2. This focal model used OLS with HC3 standard errors.

We checked whether the AI chatbot estimate depended on two plausible modeling choices: adding the original access-group assignment as a predictor and replacing each raw usage count $x$ with $\log(1+x)$. These checks address possible differences due to earlier access and the strong skew in usage, respectively.

\noindent\textbf{Mediation analysis.}
We fitted four OLS equations with intercepts to 148 students with baseline self-efficacy ($SE$), Midterm~2 ($M_2$), Midterm~3 ($M_3$), AI chatbot message count ($C$), and homework-feedback request count ($F$):
\begin{align*}
M_2 &= \alpha_1+aSE+\varepsilon_1,\\
C &= \alpha_2+d_CSE+e_CM_2+\varepsilon_2,\\
F &= \alpha_3+d_FSE+e_FM_2+\varepsilon_3,\\
M_3 &= \alpha_4+c'SE+b_MM_2+b_CC+b_FF+\varepsilon_4.
\end{align*}
Both usage counts covered the interval between Midterms~2 and 3. This path model used Midterm~2 as the intermediate achievement measure. Its Midterm~3 equation had the same predictors as the RQ2 usage--exam model, but used conventional rather than HC3 standard errors. We report standardized path coefficients and unstandardized indirect associations, calculated as products of the corresponding paths. Percentile 95\% confidence intervals used 5,000 learner-level bootstrap resamples (seed 20260606), refitting all four equations in each resample. We report approximate two-sided bootstrap $p$ values for the five indirect paths, with Holm adjustment across this family.

\section{Results}

We report the randomized access comparison first, followed by associations of use of the AI support system with exam scores and the mediation analysis.

\subsection{RQ1: Homework and Exam Scores with Earlier Access}

At baseline, the early- and later-access groups had Midterm~1 means of 85.45 ($SD=10.60$) and 83.88 ($SD=13.47$), and Homework~1--4 means of 85.09 ($SD=12.26$) and 83.59 ($SD=10.89$). While only the early-access group could use the two interfaces, 59/75 early-access students and 0/78 later-access students had at least one logged AI chatbot message or homework-feedback request. For Homework~5, raw means were 98.16 ($SD=9.20$) versus 95.77 ($SD=10.88$), and for Homework~6, means were 100.48 ($SD=10.31$) versus 95.91 ($SD=15.91$).

During the early-access period, students assigned earlier access scored 2.56 points higher on homework after adjustment for their Homework~1--4 average and homework assignment ($SE=1.22$, 95\% CI $[0.17,4.94]$, $p=.036$; 306 observations from 153 learners). This means that students assigned earlier access submitted higher-scoring homework, on average, even when prior homework performance was taken into account.

The groups did not differ significantly on Midterm~2 (raw means 69.95 for 75 earlier-access students versus 71.52 for 77 later-access students), which students completed without AI. Adjusting for Midterm~1, the early-access estimate was $B=-2.47$ (95\% CI $[-5.60,0.66]$, $p=.122$; $N=152$). These results suggest that improved AI-supported homework performance did not translate into detectable gains on the subsequent exam.

\begin{figure}[t]
\centering
\includegraphics[width=\linewidth]{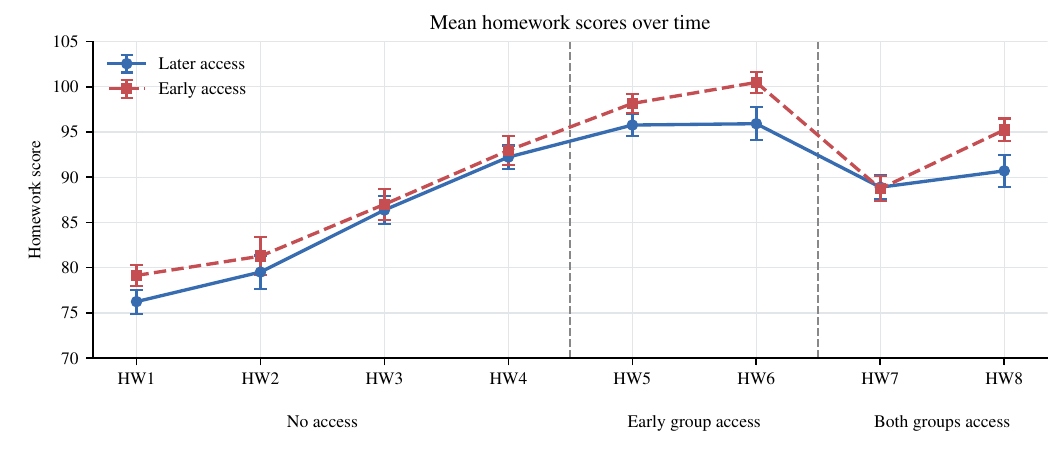}
\caption{Mean homework scores over time by randomized access group, with error bars showing $\pm1$ standard error. Dashed boundaries mark changes in access. Homework~1--6 include 75 early-access and 78 later-access students per assignment; Homework~7--8 include 74 and 76, respectively, with missing scores omitted. Scores can exceed 100 because of bonus items. Adjusted comparisons for the period when access differed are reported in the text.}
\label{fig:main_estimates}
\end{figure}

\subsection{RQ2: Student Characteristics, AI Use, and Exam Scores}

\noindent\textbf{Self-efficacy and use of the two interfaces.}

Among 152 students with self-efficacy and Midterm~2 data, 106 used the AI chatbot and 90 requested homework feedback. At the same Midterm~2 score, students whose self-efficacy scores were one point higher had predicted counts of student messages sent to the AI chatbot 39\% lower ($p<.001$) and predicted counts of homework-feedback requests 31\% lower ($p=.002$). At the same self-efficacy score, higher Midterm~2 scores were associated with lower predicted use of the two interfaces, but neither association was statistically significant: the incidence rate ratio (IRR) was .982 for student messages sent to the AI chatbot ($p=.056$) and .989 for homework-feedback requests ($p=.393$). Thus, students with lower confidence tended to use the two interfaces more often.

\Needspace{4\baselineskip}
\noindent\textbf{Interface Use and Subsequent Exam Scores.}

We next examined whether AI chatbot use between Midterms~2 and 3 was associated with Midterm~3 scores. Controlling for baseline self-efficacy, Midterm~2, and use of the homework-feedback interface, each additional student message sent to the AI chatbot was associated with a .0302-point lower Midterm~3 score (95\% CI $[-.0567,-.0036]$, $p=.026$; $N=148$), equivalent to $-3.02$ points per 100 messages in the linear model. The association for homework-feedback requests was positive but not statistically reliable ($B=0.084$, $p=.131$).

We checked whether the result changed when we accounted for earlier access or reduced the influence of very high usage counts. Adding whether students originally received early or later access gave a similar AI chatbot estimate ($B=-.0284$, 95\% CI $[-.0543,-.0026]$, $p=.031$). Using $\log(1+x)$ counts, which reduces the influence of very large counts, also produced a negative estimate ($B=-1.53$, 95\% CI $[-2.64,-.41]$, $p=.007$). The latter coefficient is on a different scale and should not be compared numerically with the per-message coefficient. The association between greater AI chatbot use and lower exam scores remained negative in both models.

Together, these results show that students who used the AI chatbot more tended to score lower on the next exam. The association was not accounted for by the measured differences in prior grades and self-efficacy.

\subsection{RQ3: Indirect Associations between Self-Efficacy and Exam Scores}

\begin{figure}[htbp]
\centering
\includegraphics[width=\linewidth]{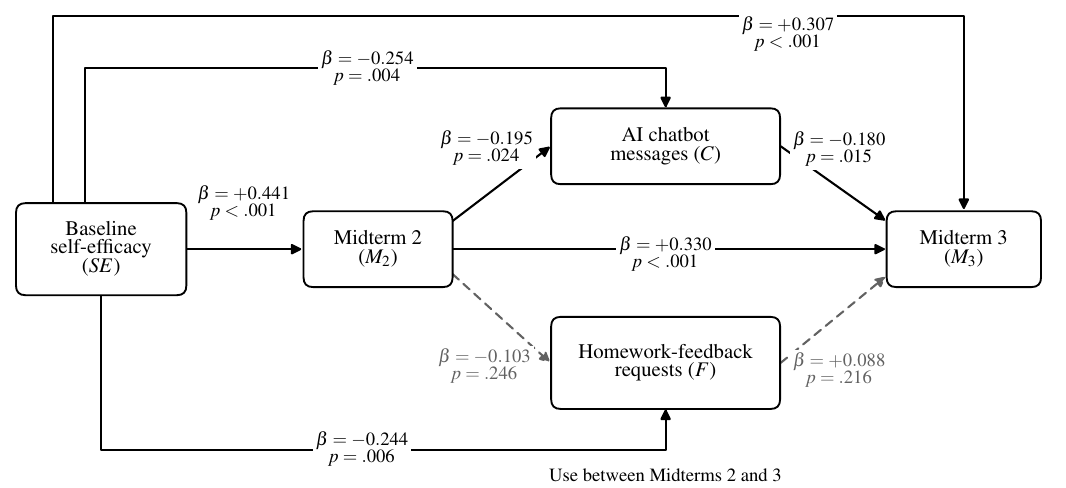}
\caption{Associations among baseline self-efficacy, prior performance, interface use between Midterms~2 and 3, and Midterm~3 scores ($N=148$). Arrows show standardized OLS coefficients ($\beta$) and model-based $p$ values; dashed paths have $p\geq .05$. The indirect association from self-efficacy through AI chatbot messages to Midterm~3 was $B=.622$ exam points per self-efficacy point (95\% percentile bootstrap CI $[.121,1.560]$; 5,000 resamples). This unstandardized indirect estimate is the product of the two raw path coefficients.}
\label{fig:mediation_paths}
\end{figure}

Figure~\ref{fig:mediation_paths} shows all fitted paths, including the direct associations with Midterm~3 and the parallel routes through the two interfaces.

Among 148 students, higher baseline self-efficacy was associated with fewer AI chatbot messages after accounting for Midterm~2 ($B=-20.62$, $\beta=-.254$, $p=.004$). In the Midterm~3 equation, more messages were associated with lower scores after accounting for self-efficacy, Midterm~2, and homework-feedback requests ($B=-.0302$, $\beta=-.180$, $p=.015$). The indirect association through AI chatbot use was $B=.622$ (95\% bootstrap CI $[.121,1.560]$; Holm-adjusted $p=.021$). Thus, lower self-efficacy was linked to greater chatbot use and, in turn, lower exam scores.

Table~\ref{tab:mediation} reports the indirect associations. The path through Midterm~2 was positive. The serial path through Midterm~2 and AI chatbot use, and both paths involving homework feedback, had intervals that included zero. The direct self-efficacy coefficient in the Midterm~3 equation was $B=4.18$ ($\beta=.307$, $p<.001$).

\begin{table}[t!]
\centering
\caption{Indirect associations between baseline self-efficacy and Midterm~3 ($N=148$). Estimates are exam points per one-point increase in self-efficacy; intervals use 5,000 percentile bootstrap resamples.}
\label{tab:mediation}
\begin{tabular}{lrrr}
\toprule
Path from self-efficacy to Midterm~3 & Estimate & 95\% CI & Holm adj.\ $p$ \\
\midrule
Through Midterm~2 & 1.975 & $[1.028,3.226]$ & $<.001$ \\
Through AI chatbot use & .622 & $[.121,1.560]$ & $.021$ \\
Through homework feedback & $-.293$ & $[-.811,.058]$ & $.246$ \\
Through Midterm~2 and AI chatbot use & .211 & $[-.029,.449]$ & $.241$ \\
Through Midterm~2 and homework feedback & $-.054$ & $[-.275,.067]$ & $.429$ \\
\bottomrule
\end{tabular}
\par\smallskip
\begin{minipage}{\linewidth}
\small Approximate two-sided bootstrap $p$ values were calculated as twice the smaller proportion of positive or negative bootstrap estimates. Adjusted $p$ values use Holm adjustment across the five indirect paths; confidence intervals are pointwise.
\end{minipage}
\end{table}

\input{interaction-cases}

\section{Discussion}

We found three main patterns. First, students randomly assigned earlier access to the two interfaces earned higher scores on homework completed while support was available, but no advantage was detected on the subsequent exam completed without AI. Second, after all students had access and chose how much to use each interface, lower self-efficacy was associated with greater use of the two interfaces, and more student messages sent to the AI chatbot were associated with lower scores on the next exam. Third, the mediation analysis identified a positive indirect association between baseline self-efficacy and subsequent exam scores through AI chatbot use: lower confidence was linked to more messages, which were associated with lower scores. The illustrative dialogues show how requests for solutions differ from requests to develop a student's own reasoning.

\subsection{Homework Performance and Independent Problem Solving}

Students assigned earlier access earned higher homework scores while AI support was available. On Midterm~2, which students completed without AI, we found no significant difference between the access groups. Because the exam assessed proof reasoning on similar types of problems but different questions, higher homework scores did not provide clear evidence of better independent performance on new proof problems.

This pattern illustrates the assistance dilemma: help that improves immediate performance does not necessarily improve students' ability to solve problems independently \cite{koedinger2007exploring}. In a mathematics field experiment, \shortciteauthor{bastani2025generative}~\citeyear{bastani2025generative} found lower scores on an assessment without AI after unguarded chatbot use; tutoring guardrails reduced this disadvantage. In a programming experiment, \shortciteauthor{bassner2026stress}~\citeyear{bassner2026stress} found higher scores during AI-supported practice but no significant advantage in knowledge gains or code comprehension. Together, these studies highlight the importance of evaluating AI-supported performance and independent understanding separately.

One interviewee distinguished reassurance about a completed solution from confidence in solving new problems: ``[\ldots] in terms of feeling more confident with new problems, not entirely, but like it makes me feel more confident like after I've done the problem that I had a correct solution.''

This account helps interpret the contrast between higher supported-homework scores and the absence of a detected exam advantage: students may use AI to check completed work without feeling better prepared to solve unfamiliar problems. Homework scores alone would have provided an incomplete account of our deployment. Evaluations should examine both the work students complete with AI and the reasoning they can perform independently.

\subsection{Interpreting the Association between Chatbot Use and Exam Scores}

Students with lower self-efficacy used the two interfaces more often, and greater AI chatbot use was associated with lower subsequent exam scores. These findings raise a question about what students do with the additional help they seek.

We interpret cognitive offloading as a likely explanation for the association between greater AI chatbot use and lower Midterm~3 scores: students may obtain proof steps without practicing how to generate them independently \cite{risko2016cognitive}. One interviewee described a perceived cost of ready access to help: ``I think, because I have access to [system name]. I actually tend to forget a lot of things.'' The student went on to explain that the system made information easy to find and gave direction on where to look for an answer. This account suggests how readily available guidance may reduce a student's own search for a solution.

Lower self-efficacy among frequent users further suggests that some students may have continued to struggle despite seeking help. The negative association also appears in the self-reported usage results of \citeauthor{lepp2025generative}~\citeyear{lepp2025generative}. Our study adds the logged number of student messages sent to the AI chatbot and the number of homework-feedback requests and examines their relationship with later exam scores while accounting for prior performance and self-efficacy.

\subsection{Student Reasoning in Chatbot Interactions}

The association between greater chatbot use and lower exam scores raises the question of whether students use AI assistance to develop their own reasoning or to obtain steps they have not worked through. The dialogue examples in Section~\ref{sec:interaction_cases} make this distinction concrete through students' requests for further solution steps and feedback on their own attempts.

In Figure~\ref{fig:chat_sequence}, increasingly direct requests elicit more of the solution. In Case A in Section~\ref{sec:interaction_cases}, the student likewise asks the chatbot to continue rather than attempting the verification it proposes. These exchanges illustrate the distinction in help-seeking research between assistance that enables learners to continue working and ineffective ways of obtaining help \cite{aleven2003help}. Students may learn from an AI-generated proof step by explaining why it follows and using that reasoning to continue the proof. Case B shows why the accuracy of the response also matters: a student offers an attempted solution, but the chatbot's incorrect diagnosis directs attention away from the error that needs to be resolved.

Interpreting what students do with an answer also requires understanding why they requested it. One student who usually solved the main homework problems before using the system recalled an exception: ``[\ldots] I asked, like a bonus problem after the fact, like a previous week's bonus problem, because I didn't really understand it. But I haven't like asked it for a main homework problem. [\ldots]'' Seeking an answer to review a previous problem differs from obtaining one in place of a current attempt, a distinction that an initial-request label alone may miss.

\subsection{Implications for Teaching and AI Design}

For instructors, these findings suggest looking beyond whether students finish an assignment or use AI frequently. Reviewing a student's attempted proof, asking them to explain a supplied step, and examining later work without AI could better reveal what they understand. A request for an answer may be an occasion to ask where the student is stuck rather than to assume misconduct or low ability; labels alone should not determine grading or disciplinary decisions \cite{slade2013ethical}.

The cases suggest different instructional responses rather than a single response to all continued dialogue. In Case A, the AI chatbot could ask the student to identify which given inverse applies before supplying another step. In Case B, it could ask whether the student's method includes a hand with two non-Jack cards of the same rank. Such prompts would make the student's next contribution informative about the specific difficulty. Systems should also check the mathematical basis of their feedback: neither withholding a complete answer nor eliciting another message ensures that the guidance is sound.

Greater chatbot use was associated with lower subsequent exam scores, whereas homework-feedback use was not negatively associated with exam performance. The homework-feedback interface required a submitted proof, making the student's current reasoning available for critique. The chatbot allowed students to request solution steps before making an attempt and to continue requesting steps instead of explaining them. The design priority is to build on students' own reasoning through feedback, rather than supply the reasoning they need to practice themselves. We therefore recommend against using an open-ended AI chatbot as a tutor and suggest exploring tutors embedded in course activities that require students to attempt and explain their reasoning before receiving further help.

To help researchers investigate how students use AI assistance in other courses, our resource package provides the exact annotation prompts, model settings, coding definitions, analysis code, and aggregate results for inspection and adaptation. The materials document how requests are labeled and how interface use is related to exam scores. This documentation helps researchers distinguish differences in student behavior from differences caused by coding and analysis choices. Researchers can adapt the prompts and code to study when answer requests support learning and when they replace students' own reasoning.

\subsection{Limitations}

The study examined AI support within one proof-oriented course, where students who consented may differ from those who did not. Generalization therefore depends on both the participating students and the instructional setting: usage patterns may differ across subjects and assessment designs, or with newer models and different prompts. Within this setting, random assignment allowed us to compare earlier and later access to the two interfaces offered together, but not to identify their separate effects. The subsequent Midterm~2 assessment captured independent performance in the ordinary course rather than performance on a purpose-built transfer or delayed-retention test.

Once all students had access, they chose whether and how often to use each interface. Unmeasured differences among students may therefore contribute to the negative usage--exam association and the indirect association linking self-efficacy to exam scores through chatbot use. Although we accounted for baseline self-efficacy and prior performance, we did not measure within-student changes in self-efficacy, time on task, external AI use, or unrecorded help. The count models also combine users and nonusers, so their estimates do not distinguish taking up an interface from using it more frequently once adopted. These features limit what the usage results can tell us about the interfaces themselves; the two usage estimates do not establish different interface effects \cite{gelman2006difference}. They also leave open how students used the assistance they received. Cognitive offloading remains an interpretation of the observed associations because mental effort and students' use of AI-generated reasoning were not directly measured.

The dialogue analysis provides a closer view of requests and responses, but its labels describe the initial request rather than how students use the reasoning that follows. Classifier agreement with human annotations was evaluated on the same 80 sessions used to develop the prompt, rather than on an independent test set. The descriptive count of students with at least one answer-seeking session also depends on how many sessions each student contributed: more sessions provide more opportunities to receive that label. The selected dialogues and interview excerpts complement these labels by illustrating how an answer request can arise and an exchange can develop, providing context for interpreting the usage associations.

\subsection{Future Work}

Future research should examine when answer-seeking helps students learn and when it replaces their own reasoning. Studies linking reflective or answer-focused dialogue to outcomes \cite{lai2025processaction,mcnichols2026studychat}, together with the sequence analysis in the StudyChat preprint \cite{rinja2026unpacking}, motivate examining what follows a help request. Complete conversations could show whether a student asks why a step works, revises an attempted proof, or simply moves on after requesting an answer. Combining these conversations with students' written work and later performance without AI would help distinguish productive use of an answer from dependence on it.

The prepared annotation prompts and analysis code provide a starting point for studying these patterns across more students, courses, and AI tools. Researchers could examine whether the negative associations observed here recur in other settings and whether the associations vary by task, students' prior knowledge, or the form of assistance. Applying the prompts to complete conversations may require revised definitions, and labels should be checked against independent human coding in each new setting. These larger studies should account for how often students use the tools, rather than treating a single answer-seeking request as equivalent across students with very different amounts of use.

A future randomized study could compare three arms: chat that provides answers, chat that provides hints without answers, and feedback on submitted proof attempts, while holding the underlying model, course content, and other features constant wherever possible. The study should examine both immediate task performance and later performance on problems completed without AI. Further comparisons could test when to offer a hint or ask the student to attempt or explain a step before receiving more help. The goal is to identify assistance that helps students complete their work and develop independent reasoning.

\Needspace{13\baselineskip}
\section{Conclusion}

Students assigned earlier access to the two interfaces earned higher homework scores, but we found no significant difference between access groups on the next exam, which students completed without AI. During later self-selected use, students with lower self-efficacy used both interfaces more often, and greater AI chatbot use was associated with lower subsequent exam scores.

The mediation analysis linked lower baseline self-efficacy to greater AI chatbot use and, in turn, lower exam scores. The dialogue examples in Section~\ref{sec:interaction_cases} complement this pattern by showing how students may obtain proof steps without practicing the reasoning needed to produce them independently. We therefore recommend against using an open-ended chatbot as a tutor and suggest exploring course-embedded tools that provide feedback on students' own proof attempts. AI-supported homework, including work completed with these tools, should be evaluated by whether students can subsequently construct and explain proofs without assistance.

\section*{Conflict of Interest}
The authors declare no conflict of interest.

\section*{Ethics Statement}
The study received institutional review board approval. Participation was voluntary and did not affect course grades. Students consented to research use of course, survey, and AI support system records under the approved protocol.

\section*{Data Availability Statement}
Aggregate results and supporting code and prompts are available at \url{https://anonymous.4open.science/r/lak27-anon}. Student-level course records, interaction logs, and interview records are not included; access requires applicable institutional and protocol approval.

\section*{AI-Use Disclosure}
GPT-5.2 (with the snapshot reported in the Methods) was used at temperature 0 as a post hoc classifier of study sessions. The authors initially drafted the manuscript. Hermes Agent, powered by OpenAI GPT-5.6 Sol, subsequently assisted with language editing, analysis-code review, and consistency checking. The authors take full responsibility for the final manuscript.

\phantomsection
\bibliography{main}
\end{document}

%% file: interaction-cases.tex
\Needspace{10\baselineskip}
\subsection{Illustrative Student--AI Chatbot Interactions}\label{sec:interaction_cases}

Each session has four turns, excerpted in order below. Bracketed ellipses mark omissions, and mathematical formatting is adapted. Three coding dimensions describe the first three messages: initial-request intent, response completeness, and student follow-up. Answer-seeking and help-seeking describe the initial student request (message 0). Partial answer/hint describes the AI chatbot's response (message 1), with response completeness coded only after an answer-seeking request. Engage describes the student's next message (message 2), including a reply such as ``I am not sure''; it indicates continuation of the exchange rather than demonstrated learning. Message 3 was not provided to the classifier and remains unlabeled.

\noindent\textbf{Case A: Asking the AI chatbot to continue a solution.}
The problem asks students to find $x$ satisfying $x\equiv a\pmod n$ and $x\equiv b\pmod m$, where $m$ and $n$ are relatively prime. The given inverses satisfy $mr\equiv1\pmod n$ and $ns\equiv1\pmod m$.
\begin{quote}
\textbf{0. Student (answer-seeking):} ``[\ldots] Can you solve this? I'm going to try to make sense of your solution''

\textbf{1. Tutor (partial answer/hint):} ``[\ldots] By the Chinese Remainder Theorem, such a solution $x$ can be constructed as: $x = a \cdot m \cdot s + b \cdot n \cdot r$ [\ldots] Can you explore whether these congruences hold true using the above representation of $x$? What would happen if you substitute and simplify?''

\textbf{2. Student (engage):} ``Could you do the next step''

\textbf{3. Tutor (unlabeled):} ``[\ldots] Given $m \cdot s \equiv 1 \pmod{n}$, how does this help with simplifying $a \cdot m \cdot s \pmod{n}$? [\ldots]''
\end{quote}
The student expresses a wish to understand the solution but asks the AI chatbot to carry out the next step. When invited to verify the formula, the student continues the conversation without attempting that verification. The exchange thus shows how a stated learning goal can coexist with delegating the reasoning needed to reach it.

The AI chatbot swaps the two inverses: the construction should be $x=amr+bns$. In the final response, it also assumes $ms\equiv1\pmod n$, although this relation is not given.

\noindent\textbf{Case B: Seeking help with an attempted solution.}
\begin{quote}
\textbf{0. Student (help-seeking):} ``[\ldots] How many 5 card hands can you form with exactly 2 Jacks? [\ldots] My first step was also to choose 2 jacks out of four in a deck (4 choose 2) My second step however, was to choose ranks for all the other cards (12 choose 3) [\ldots] My third step was to choose suits for each of them (4 choose 1)$^{3}$''

\textbf{1. Tutor (answer level not applicable):} ``[\ldots] How might fixing the rank and suit combinations in your approach create duplicates or omissions compared to simply picking 3 cards out of 48? [\ldots]''

\textbf{2. Student (engage):} ``I am not sure''

\textbf{3. Tutor (unlabeled):} ``[\ldots] this method might lead to overcounting combinations --- because different ranks can have the same suit choice, or same rank cards could be combined in a different suit arrangement. [\ldots]''
\end{quote}
The student shares an attempted solution, but the AI chatbot incorrectly describes the problem as overcounting. Choosing three distinct non-Jack ranks excludes valid hands in which the remaining cards share a rank; it does not count the same hand more than once.

Unlike Case A, the student has already made the reasoning available for diagnosis. The full initial request also gives the accepted answer, so the difficulty concerns why the alternative method fails, not simply obtaining a result. Asking the student a question is therefore not sufficient: the question must help examine the actual error in the student's reasoning. Neither session includes a later student explanation that resolves the difficulty.